%% file: paper.tex
\documentclass[conference, a4paper]{IEEEtran}
\IEEEoverridecommandlockouts
\usepackage{svg}
\usepackage{multirow}
\usepackage{amsmath,amssymb,amsfonts,mathtools}
\usepackage{placeins}
\usepackage{algorithm,algpseudocode}
\usepackage{orcidlink} 
\usepackage{listings}
\usepackage{tikz}
\usetikzlibrary{shapes.geometric, arrows}

\usepackage{graphicx}
\usepackage{balance}
\usepackage{colortbl}

\usepackage[capitalise, noabbrev]{cleveref}
\crefname{lstlisting}{listing}{listings} 
\Crefname{lstlisting}{Listing}{Listings} 

\usepackage[nolist]{acronym}
\usepackage{xcolor}
\definecolor{lightergray}{rgb}{0.95, 0.95, 0.95} 
\definecolor{darkgreen}{rgb}{0.1, 0.8, 0.6} 
\usepackage{circledsteps}

\usepackage{siunitx}
\usepackage{pifont}
\usepackage{subcaption}
\usepackage{graphicx}
\algrenewcommand\algorithmiccomment[1]{\hfill /*#1*/}

\def\BibTeX{{\rm B\kern-.05em{\sc i\kern-.025em b}\kern-.08em
    T\kern-.1667em\lower.7ex\hbox{E}\kern-.125emX}}

\bstctlcite{IEEEexample:BSTcontrol}

\makeatletter
\newcounter{problem}
\newcounter{tmp}

\makeatother

\usepackage[all]{background}
\usepackage{stackengine}
\setstackEOL{\\}
\setstackgap{L}{\normalbaselineskip}
\SetBgContents{\color{gray}{\tiny \Longstack{
    PREPRINT - Published at the 62th Design Automation Conference (DAC), DOI: 10.1109/DAC63849.2025.11133237\\
    \textbf{© 2025 IEEE:} Personal use of this material is permitted. Permission from IEEE must be obtained for all other uses, \\ including reprinting/republishing this material for advertising or promotional purposes, collecting new collected works for resale \\ or redistribution to servers or lists, or reuse of any copyrighted component of this work in other works.
}}}
\SetBgPosition{4.5cm,1cm}
\SetBgOpacity{1.0}
\SetBgAngle{0}
\SetBgScale{1.8}

\begin{document}
\bstctlcite{IEEEexample:BSTcontrol}

\input{abbrev.tex}

\title{
Introducing Instruction-Accurate Simulators for Performance Estimation of Autotuning Workloads
}

\def\finalpaper{1}

\if\finalpaper1
{
    \author{
    \IEEEauthorblockN
        {Rebecca Pelke\orcidlink{0000-0001-5156-7072},
        Nils Bosbach\orcidlink{0000-0002-2284-949X},
        Lennart M. Reimann\orcidlink{0009-0003-5825-2665},
        Rainer Leupers\orcidlink{0000-0002-6735-3033}}
    \IEEEauthorblockA{\textit{Institute for Communication Technologies and Embedded Systems}\\
        \textit{RWTH Aachen University, Germany}\\
        \{pelke, bosbach, lennart.reimann, leupers\}@ice.rwth-aachen.de}
    }
}
\else
    \author{
      \IEEEauthorblockN{Authors are removed for submission version}
      \\
      \\
      \IEEEauthorblockA{Affiliations are removed for submission version}
      \vspace{-0.5cm}
      \\
    }
\fi

\maketitle

\begin{abstract}
Accelerating \ac{ml} workloads requires efficient methods due to their large optimization space.
Autotuning has emerged as an effective approach for systematically evaluating variations of implementations.
Traditionally, autotuning requires the workloads to be executed on the target \ac{hw}.
We present an interface that allows executing autotuning workloads on simulators.
This approach offers high scalability when the availability of the target \ac{hw} is limited, as many simulations can be run in parallel on any accessible \ac{hw}.


Additionally, we evaluate the feasibility of using fast instruction-accurate simulators for autotuning.
We train various predictors to forecast the performance of \ac{ml} workload implementations on the target \ac{hw} based on simulation statistics.

Our results demonstrate that the tuned predictors are highly effective.
The best workload implementation in terms of actual run time on the target \ac{hw} is always within the top \SI{3}{\percent} of predictions for the tested x86, ARM, and RISC-V-based architectures.
In the best case, this approach outperforms native execution on the target \ac{hw} for embedded architectures when running as few as three samples on three simulators in parallel.

\end{abstract}


\acresetall
\acused{cpu}

\input{introduction.tex}
\input{background.tex}
\input{implementation.tex}
\input{results.tex}

\vspace{0.4cm}
\section{Conclusion and Future Work}
\label{sec:conclusion}

\vspace{0.1cm}
In this work, we introduced an interface for executing autotuning workloads on simulators and explored the feasibility of using instruction-accurate simulators for autotuning of \ac{ml} workloads.
We trained and compared different score predictors for the x86, ARM, and RISC-V architectures.
Our results show that the tuned predictors can identify optimal implementations within the top \SI{3}{\percent} of predictions.
In the case of limited availability of the target \ac{hw}, our approach can even be used to accelerate autotuning.
In the best case, 3 parallel simulations on the used x86 machine were sufficient to replace one RISC-V board.
Our research lays the groundwork for using instruction-accurate simulations for performance estimation.

Future work will focus on benchmarking a broader range of \acp{cpu} to train more generalized predictors.
These generalized predictors can then be applied to previously untested CPUs,
enhancing our methodology's appeal for pre-silicon software development.

\balance
\bibliographystyle{IEEEtran}
\bibliography{bibtexentry}

\end{document}

%% file: abbrev.tex
\begin{acronym}
    \acro{gemm}[GEMM]{General Matrix Multiply}
    \acro{nn}[NN]{Neural Network}
    \acro{dnn}[DNN]{Deep Neural Network}
    \acro{mvm}[MVM]{Matrix Vector Multiplication}
    \acro{mmm}[MMM]{Matrix Matrix Multiplication}
    \acro{tvm}[TVM]{Tensor Virtual Machine}
    \acro{ml}[ML]{Machine Learning}
    \acro{te}[TE]{Tensor Expression}
    \acro{tir}[TIR]{Tensor Intermediate Representation}
    \acro{dag}[DAG]{Directed Acyclic Graph}
    \acro{cpu}[CPU]{Central Processing Unit}
    \acro{l1i}[L1I]{L1 Instruction}
    \acro{l1d}[L1D]{L1 Data}
    \acro{llc}[LLC]{Last Level Cache}
    \acro{os}[OS]{Operating System}
    \acro{iss}[ISS]{Instruction-Accurate Simulator}
    \acro{fss}[FSS]{Full System Simulator}
    \acro{simd}[SIMD]{Single Instruction Multiple Data}
    \acro{mlr}[MLR]{Multiple Linear Regression}
    \acro{mae}[MAE]{Mean Absolute Error}
    \acro{mse}[MSE]{Mean Squared Error}
    \acro{rss}[RSS]{Residual Sum of Squares}
    \acro{hw}[HW]{hardware}
    \acro{sw}[SW]{software}
\end{acronym}

%% file: introduction.tex
\section{Introduction}
\label{sec:intro}

The optimization of \ac{ml} models is crucial due to their high computational demands.
Traditional analytical methods often fall short given the huge search space for optimal implementations on modern \ac{cpu} architectures.
To address this challenge, autotuning has emerged as a powerful approach.
Autotuning systematically evaluates multiple implementations of the same workload using mathematical models or \ac{ml} techniques to guide subsequent selections of implementations~\cite{datta2008stencil,lin2009auto}.
Apache \ac{tvm}~\cite{chen2018tvm}, a well-known \ac{ml} compiler framework,
implements several autotuning strategies~\cite{zheng2020ansor,hsieh2023dloopt}.
Autotuning typically requires execution on real hardware, which introduces non-determinism due to factors such as the system load~\cite{grunewald1996towards},
cache collisions~\cite{zhuravlev2012survey}, thermal throttling~\cite{benoit2020impact}, and frequency and voltage scaling~\cite{brodowski2013cpu}.

\begin{figure}[!tp]
    \centering
    \includegraphics[width=.95\linewidth, keepaspectratio]{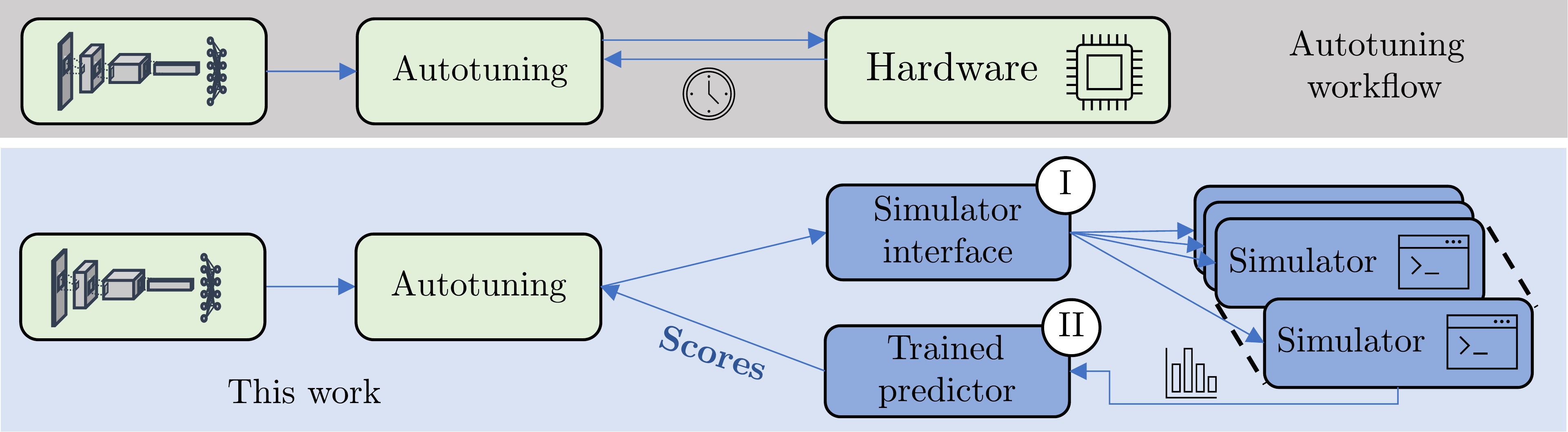}
    \caption{The proposed simulator interface \Circled{I} and the score predictor approach using instruction-accurate simulators \Circled{II}}
    \label{fig:idea}
    \vspace{-0.6cm}
\end{figure}

To mitigate these issues, each benchmark is executed multiple times, outliers are removed, cooldown periods are inserted, and caches are flushed before each repetition.
Consequently, benchmarking a single implementation takes significantly longer than its actual run time.
This is time-consuming, especially when only limited \ac{hw} devices are available.
This paper presents the following two main contributions:

\textbf{Contribution \Circled{I}: Simulator Interface - }
We present an interface allowing autotuning workloads to be executed on simulators rather than real hardware.
\cref{fig:idea} \Circled{I} illustrates this approach.
We extract the \ac{tvm} tasks and provide them as executables for the target architecture through a simulator interface.
The number of parallel tasks is configurable, each running in a separate instance of the simulator.
Potential scenarios that profit from our \textbf{simulator interface} are:

\begin{itemize}
\item The \ac{hw} is not yet available, e.g., for pre-silicon \ac{sw} development.
\item The embedded \ac{hw} is only available in limited quantities, making parallel execution of simulations on high-performance computers faster than native execution.
\item Other metrics besides run time should be optimized.
\end{itemize}

\textbf{Contribution \Circled{II}: Score Predictor - }
We aim to demonstrate that instruction-accurate simulators can be used for performance analysis, using autotuning workloads as an example.
\textit{To the best of our knowledge, we are the first to show how performance analysis can be conducted with fast instruction-accurate simulators.}
Many open-source implementations of instruction-accurate simulators exist for different architectures, e.g., QEMU~\cite{bellard2005qemu} or gem5~\cite{binkert2011gem5}.
\cref{fig:idea} \Circled{II} illustrates this idea.
A predictor gets statistics from an instruction-accurate simulator as input and calculates a score based on reference values measured on real hardware.
The score is then returned to the autotuning framework.

Since an instruction-accurate simulator does not provide accurate timing, we do \textbf{not} aim to predict execution latencies.
Instead, our approach utilizes scores to evaluate and compare different implementations of the same workload.
These scores are essential for guiding the autotuning process but are not suitable for comparing different types of workloads.
We employ multiple predictors for this task: \ac{mlr}~\cite{montgomery2021introduction},
\acp{dnn}~\cite{du2017hierarchical}, Bayesian optimization~\cite{frazier2018bayesian}, and XGBoost~\cite{chen2015xgboost}.
We tune and compare them to identify the most accurate predictions for common \ac{ml} kernels.
We evaluate our approach on different \ac{cpu} architectures, namely x86, ARM, and RISC-V.


%% file: background.tex
\section{Background and Related Work}
\label{sec:background}

\subsection{Autotuning in TVM}
\label{sec:autotuning}
Apache \ac{tvm}~\cite{chen2018tvm} is an open-source \ac{ml} compiler framework designed to optimize computations across various hardware platforms.
In \ac{tvm}, each operation, called \textit{kernel} in this work, can be expressed in different abstractions, e.g., in \ac{te} or in \ac{tir}.
For example, a kernel can be a \ac{nn} layer.
\ac{tvm} enables the application of \textit{transformation primitives} to optimize kernels for diverse target architectures.

Similar to Halide~\cite{ragan2013halide}, \ac{tvm} distinguishes between the \textit{compute} operation,
which defines the functional behavior of the kernel, and the \textit{schedule}, which defines its implementation.
This distinction allows a single computation to have multiple schedules.
The set of all possible schedules of an operation is called \textit{design space}.
Large design spaces and complex hardware behavior make analytical approaches impractical for finding optimal schedules for \ac{ml} kernels.
Auto-tuning addresses this issues by empirically evaluating multiple schedules directly on the target hardware.
\ac{tvm} offers three autotuning concepts:
the AutoTVM framework, the Auto-Scheduler (also called Ansor~\cite{zheng2020ansor}), and the Meta-Scheduler.
This work focuses on AutoTVM and the Auto-Scheduler since they operate on the same input representation, namely \ac{te}.

\Cref{lst:TEMatMulOperation} provides an example of a \ac{te} compute operation definition.
Tensor $C$ (size ${N\times M}$) contains the result of a \ac{mmm} between matrices $A$ (size ${N\times L}$) and $B$ (size ${L\times M}$).

\begin{figure}[!hbp]
\vspace{-0.4cm}
\centering
\begin{lstlisting}[language=Python, xleftmargin=0.3cm,
    caption=Definition of a \ac{mmm} compute operation in \ac{te},
    label=lst:TEMatMulOperation
]
k = te.reduce_axis((0, L), name="k")
C = te.compute((N, M), lambda i, j: \
    te.sum(A[i,k]*B[k,j], axis=k), name="matmul")
\end{lstlisting}
\vspace{-0.5cm}
\end{figure}

AutoTVM requires users to define templates with tunable parameters, such as loop tiling factors.
AutoTVM then navigates this search space by executing configurations on real hardware and measuring the run time.
It requires some expertise to define a useful schedule template.
However, pre-designed templates for common operators can be found in the repository.
As an example, \cref{lst:AuotTVMSearchSpaceDefinition} illustrates how a \textit{split} primitive of a single axis can be described using AutoTVM.

\begin{figure}[!htp]
\vspace{-0.4cm}
\centering
\begin{lstlisting}[language=Python, xleftmargin=0.3cm,
    caption=Definition of a scheduling template for AutoTVM,
    label=lst:AuotTVMSearchSpaceDefinition,
    numbers=left
]
s = te.create_schedule(C.op)
y,_ = s[C].op.axis
# Define schedule template
cfg = autotvm.get_config()
cfg.define_split("split_y", y, num_outputs=2,...)
# Apply schedule
yo, yi = cfg["split_y"].apply(s, C, y)
\end{lstlisting}
\vspace{-0.4cm}
\end{figure}

In contrast to AutoTVM, the Auto-Scheduler automates the schedule generation process without requiring manual template definitions,
making it more suitable for non-experts.

\begin{figure}[!tp]
    \centering
    \includegraphics[width=.85\linewidth, keepaspectratio]{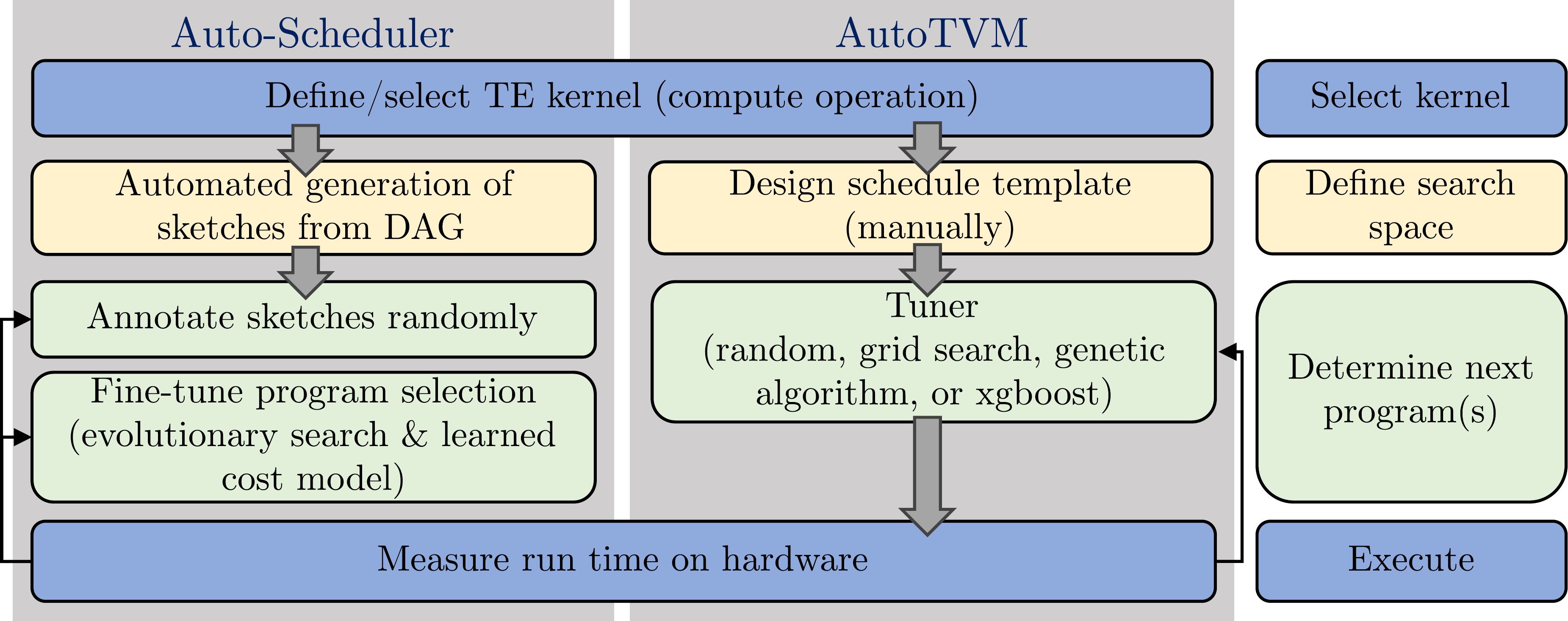}
    \caption{Autotuning using Auto-Scheduler and AutoTVM}
    \label{fig:autotuning}
    \vspace{-0.3cm}
\end{figure}

\cref{fig:autotuning} illustrates the workflows of both the Auto-Scheduler and AutoTVM.
Unlike the manually defined search space in AutoTVM, the Auto-Scheduler uses \textit{sketches} generated from the kernel’s \ac{dag} through predefined derivation rules.
A sketch basically contains nested loops with placeholders, which are filled during an \textit{annotation} phase.
During annotation, loop axes can also be marked for parallelization, unrolling, or vectorization.
Implementation details can be found in~\cite{zheng2020ansor}.
In contrast to the Auto-Scheduler, AutoTVM relies on tuners responsible for selecting subsequent programs based on selectable tuning algorithms.

\subsection{Cache Hierarchy}
\label{sec:caches}
\begin{figure}[!tp]
    \centering
    \includegraphics[width=.9\linewidth, keepaspectratio]{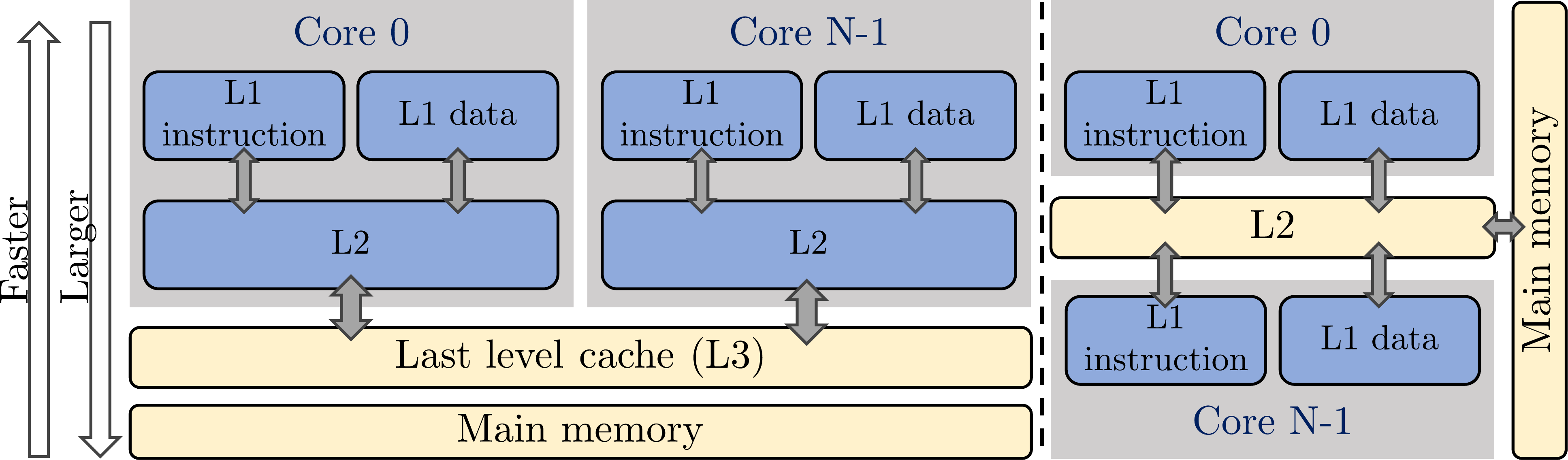}
    \caption{Typical cache hierarchies of modern \acp{cpu}}
    \label{fig:cachehierarchies}
    \vspace{-0.5cm}
\end{figure}

\ac{ml} operations require cache optimizations to achieve high performance~\cite{kelefouras2023design,li2021analytical}.
Modern \acp{cpu} have hierarchical memory systems with multiple cache levels (L1, L2, L3) as shown in \cref{fig:cachehierarchies}.
The hierarchy can vary between different \acp{cpu}~\cite{kumar2016overview}.
The L1 cache is usually divided into Data (L1D) and Instruction (L1I) cache.
Higher-level caches are often shared among the cores.
Most \acp{cpu} use $N$-way set-associative caches, where each memory address maps to one of $N$ \textit{ways} within a specific \textit{set}.
In Linux environments, information on the cache hierarchy can be accessed via the \texttt{sysfs}.

\subsection{The gem5 Simulator}
\label{sec:gem5background}
gem5~\cite{binkert2011gem5} is an open-source \ac{fss} used in computer architecture research, supporting different architectures like x86, ARM, and RISC-V.
It provides different abstraction levels.
In gem5's \textit{atomic} mode, memory accesses are executed within a single transaction.
The requester of the memory access is blocked until the access is completed.
In the \textit{timing} mode, gem5 provides detailed simulation of memory access timing, including latency, queuing delays, and bandwidth constraints.
gem5 also offers different \ac{cpu} models.
The \texttt{SimpleCPU} does not model a pipeline, which makes it fast but not very accurate in terms of timing.
It can be used in both, atomic and timing mode.
gem5 further provides an \texttt{InOrderCPU} and \texttt{O3CPU} (out-of-order \ac{cpu}), which are both equipped with a \ac{cpu} pipeline model.
In addition, gem5 distinguishes between the \textit{full-system} mode and the \textit{system call emulation} mode.
The system call emulation mode aims to simulate user-space programs.
It intercepts system calls of the target software and handles them on the host. 
The system call emulation mode is faster than the full-system mode and focuses on software testing, as it does not simulate the \ac{os} and thus cannot capture any \ac{os}-specific behavior.
\vspace{-0.3cm}

%% file: implementation.tex
\section{Implementation}
\label{sec:implementation}

In the first part of this chapter, we explain the implementation of the simulator interface.
In the second part, we use this interface to connect \ac{tvm}'s autotuning with an x86, an ARM, and a RISC-V-based instruction-accurate simulator.

\subsection{Simulator Interface}

The simulator interface allows for executing autotuning implementations on user-level or syscall-emulation simulators instead of real hardware.
This enables parallel execution of different implementations.
TVM's autotuning requires a \textit{builder} and a \textit{runner}.
The builder generates an object file that contains the compiled functionality of the workload.
The runner executes the workload using the \texttt{tvm::runtime}.

For a simulator, a standalone executable is needed.
The executable prepares the input tensors, allocates space for the output tensors, and calls the compiled workload.
To integrate this behavior into AutoTVM, we implement a custom runner called \texttt{SimulatorRunner} by inheriting from the \texttt{Runner} class (see \cref{lst:runnerAutoTVM}).
When the \texttt{run} function of the custom runner is called, it generates a main function, compiles it, and links it against the compiled object file.
A function called \texttt{simulator\_run} is called with the path to the executable.
This function serves as a simulator interface and can be overwritten to use a simulator for execution.
The return value needs to be a score quantifying the performance of the workload, e.g., the run time.
A parameter named \texttt{n\_parallel} defines how many simulators can be instantiated in parallel.

\begin{figure}[!tbp]
    \centering
    \begin{lstlisting}[language=Python, numbers=left, xleftmargin=0.3cm,
        caption=Custom run function for AutoTVM flow,
        label=lst:runnerAutoTVM
    ]
class SimulatorRunner(Runner):
    def __init__(self, n_parallel=16, ...) :
        super(SimulatorRunner, self).__init__(...)
    def run(self, measure_inputs, build_results):
        # Build executables
        # Run executables in parallel on simulator
        return [AutoTVMRes0, AutoTVMRes1, ...]
    \end{lstlisting}
    \vspace{-0.8cm}
\end{figure}

To integrate a simulator into the Auto-Scheduler, we override a function in TVM's function registry called \texttt{auto\_scheduler.local\_runner.run}.
This can be seen in \cref{lst:runnerAutoScheduler}.
The return value is a list of \texttt{AutoSchedResults} containing scores.
The remaining implementation works in the same way as for AutoTVM.

\begin{figure}[!tbp]
    \centering
    \begin{lstlisting}[language=Python, numbers=left, xleftmargin=0.3cm,
        caption=Custom run function for Auto-Scheduler flow,
        label=lst:runnerAutoScheduler
    ]
@tvm._ffi.register_func(<func_name>, override=True)
def local_run(inputs, build_results, ...) :
    # Build executables
    # Run executables in parallel on simulator
    return [AutoSchedRes0, AutoSchedRes1, ...]
    \end{lstlisting}
    \vspace{-0.8cm}
\end{figure}

\begin{figure}[!tbp]
\centering
\begin{lstlisting}[language=Python, numbers=right, xrightmargin=0.3cm,
    caption=Conv2D+Bias+ReLU kernel definition in TVM,
    label=lst:conv2d
]
def conv2d(N,H,W,CO,CI,KH,KW,...)
    ifm=te.placeholder((N,CI,H,W),...)
    weights=te.placeholder((CO,CI,KH,KW),...)
    bias=te.placeholder((N,CO,1,1),...)
    conv=topi.nn.conv2d_nchw(ifm,weights,...)
    ofm=topi.nn.relu(conv+bias)
    # Return values: transferred as DLPack tensors
    return [ifm,kernel,bias,ofm]
\end{lstlisting}
\vspace{-0.8cm}
\end{figure}

When generating an object file for a \ac{ml} kernel, the kernel is optimized and lowered through \ac{tvm}.
LLVM is used for code generation.
To produce cross-compiled object files, it is possible to specify an LLVM triple in \ac{tvm}.

\cref{lst:conv2d} shows the definition of a Conv2D+Bias+ReLU kernel as an example.
The tensor shapes and parameters are passed as command line arguments to the executable.

\subsection{gem5 Simulator Setup}
\label{sec:gem5setup}
To accurately predict timing, a cycle-accurate simulator is needed.
However, cycle-accurate simulators often suffer from slow simulation speed.
Additionally, obtaining cycle-accurate simulators with the required level of detail for commercial architectures is a challenge,
as information about precise latencies associated with, e.g., memory components, buses, or microarchitecture details are rarely available open source.

\textit{
We will demonstrate the feasibility of using a non-timing-accurate simulator for effective predictions based on quantitative parameters,
without relying on timing information.
}

Our focus is on single-core workloads.
To achieve fast simulation performance, we employ gem5 in \textit{atomic} mode with the \texttt{SimpleCPU} model for x86, RISC-V, and ARM architectures (see \cref{sec:gem5background}). 
gem5 allows to replicate the cache architecture while making it parameterizable to account for cache hits, misses, and replacements within the memory hierarchy (see \cref{sec:caches}).

\begin{figure}[!bp]
    \vspace{-0.4cm}
    \centering
    \includegraphics[width=.9\linewidth, keepaspectratio]{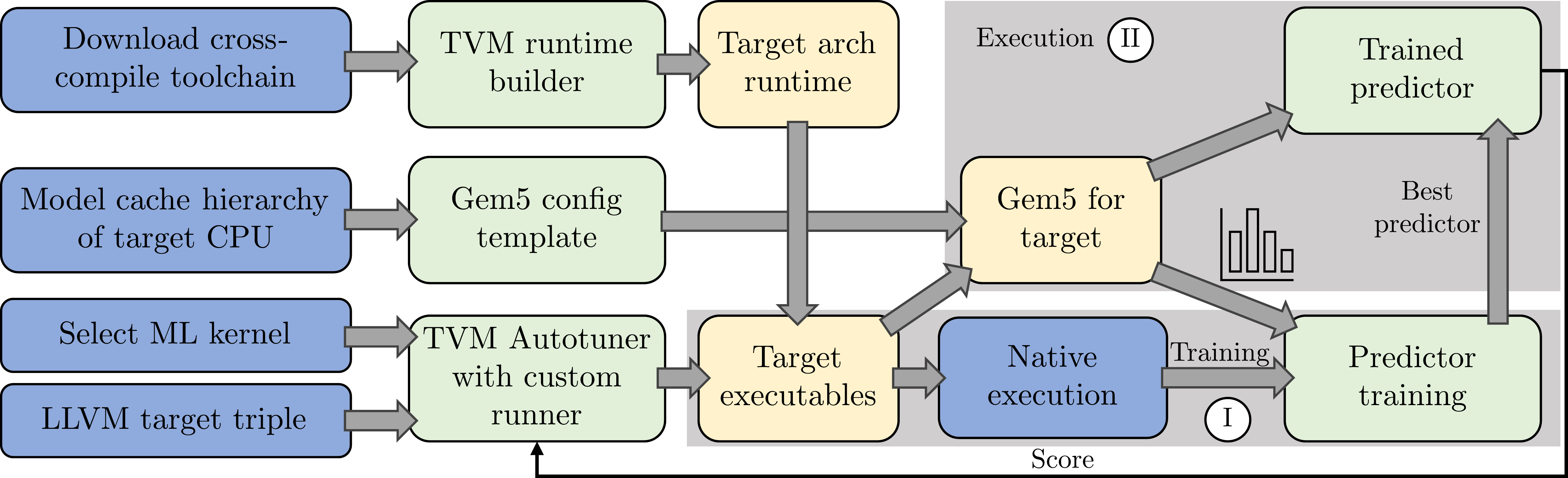}
    \caption{Workflow of training \Circled{I} and execution \Circled{II} of a predictor for one target architecture and one kernel type}
    \label{fig:toolworkflow}
\end{figure}

\subsection{Score Predictor - Workflow and Notations}
\label{sec:notations}
To use non-timing-accurate, i.e., instruction-accurate simulators, for score prediction, a predictor must be trained.
\cref{fig:toolworkflow} illustrates the training and execution phases of this predictor.
During the training phase, workloads are not only executed in gem5 but also natively on the target CPU.
We require a distinct predictor for each architecture and \textit{kernel type}.
For instance, Conv2D+Bias+ReLU represents one specific kernel type.
Note that the corresponding predictor can be applied to any combination of shapes and parameters of this kernel type.
A fixed combination of shapes and parameters for a given kernel type is referred to as a \textit{group}.
The autotuning process generates various \textit{implementations} (schedules) for each group.
In the subsequent execution phase, we leverage the pre-trained predictor.
The target CPU is \textbf{not} required anymore at this stage,
which enables the simulation of architectures such as RISC-V on x86 platforms.

\subsection{Score Predictor Training}
\label{sec:predictortraining}
The score predictor estimates a score ${S}$ for an implementation ${I}$.
The underlying principle is that these scores correlate with the run times, but only for different implementations of the same kernel type and group.
Perfect prediction means ${S_{I1} < S_{I2} \Rightarrow t_{I1} < t_{I2}}$,
where ${t_{Ix}}$ is the run time of implementation ${x}$.
Notably, comparisons between implementations across different kernel types or groups are not possible using the score.
Given that timing details are unavailable, the score predictor relies solely on quantitative information rather than latencies.
The relevant statistics derived from gem5 are:
\begin{itemize}
\item The number of the executed load/store/branch instructions divided by the total number of instructions.
\item The total number of the executed instructions normalized to the total number of executed instructions of the group.
\item Cache read/write replacements/hits/misses divided by read/write accesses of each cache.
\end{itemize}


Considering, e.g., L1D write (${L1Dw}$) cache hits for implementation ${x}$, we determine:
\vspace{-0.3cm}
\begin{align}
    P_{L1Dw}(I_x)=\frac{L1D_{w,hits}(I_x)}{L1D_{w,access}(I_x)}
\end{align}
Additionally, all parameters are also normalized to the group:
\begin{align}
    P_{L1Dw,norm}(I_x)=\frac{P_{L1Dw}(I_x)-\overline{P_{L1Dw}(\mathbf{I})}}{\overline{P_{L1Dw}(\mathbf{I})}} \label{eq:normalize}
\end{align}
The values ${\overline{P_{L1Dw}(\mathbf{I})}}$ represent the mean values across the group.
We found that the most promising approach is to use these parameters as inputs for the predictor in both their original form $P_{L1Dw}(I_x)$ and their normalized form $P_{L1Dw,norm}(I_x)$.
The output scores that are used for training are the measured run times normalized to the group, similar to \cref{eq:normalize}.
Based on these inputs and the scores, we train different predictors, which can be used with different loss functions.
In this work, we use \ac{mse}, \ac{mae}, and \ac{rss}.
Below, a brief overview of the predictors is provided.




\subsubsection{Multiple Linear Regression}
\ac{mlr} is an extension of linear regression.
\ac{mlr} is a simple predictor. It only models linear relationships.
It finds a relation between multiple outputs ${y_k}$ and multiple independent inputs ${x_i}$~\cite{uyanik2013study}, in our case:
\begin{align}
y = b_0 + b_1x_1 + b_2x_2 + ... + b_nx_n 
\end{align}

\subsubsection{Regression with DNNs}
Regression using \acp{dnn} means predicting continuous outputs based on several input features.
Typically, the networks consist of multiple fully connected  layers, with the number of input neurons corresponding to the number of input features.
Each hidden layer usually applies an activation function to introduce non-linearity and allow the network to capture complex relationships.
The final layer has a number of neurons equal to the number of output variables.
A common architecture might include a couple of hidden layers with decreasing numbers of neurons.

\subsubsection{Bayesian Optimization}
Bayesian optimization builds a surrogate model (in this case, a Gaussian process) that approximates the real/true objective function.
It uses this model to predict which areas of the parameter space are likely to yield the best results and evaluates the function in those regions.
An acquisition function balances the trade-off between exploring new regions of the space (exploration) and refining promising known regions (exploitation)~\cite{frazier2018tutorial}.
\cref{lst:bayesFunction} shows the choice of the used objective function and loss function.

\begin{figure}[!hbp]
    \vspace{-0.4cm}
    \centering
    \begin{lstlisting}[language=Python, numbers=right, xrightmargin=0.3cm,
        caption=Objective function for bayes optimization,
        label=lst:bayesFunction
    ]
def objective_function(C, RBF_scale, noise):
    func = ConstantKernel(constant_value=C, ...) \
        * RBF(length_scale=RBF_scale, ...) \
        + WhiteKernel(noise_level=noise, ...)
    gp = GaussianProcessRegressor(func) \ 
        .fit(X_train, y_train)
    predictions = gp.predict(X_test)
    return - loss_function(y_test, predictions)
    \end{lstlisting}
    \vspace{-0.5cm}
\end{figure}

The hyperparameters of the Gaussian process are the input parameters of the \texttt{objective\_function}.
The Bayesian optimization framework maximizes the objective function through a series of iterations.
Each iteration involves fitting a Gaussian process model with new hyperparameters and updating the probabilistic model based on the results.

\subsubsection{XGBoost}
Extreme Gradient Boosting (XGBoost~\cite{chen2019package}) is a powerful algorithm for regression tasks.
It uses an ensemble of decision trees to predict continuous outcomes.
In XGBoost, trees are built sequentially, with each tree trying to minimize the errors made by the previous ones by adjusting residuals.
The algorithm optimizes a loss function using gradient descent.
The hyperparameters of the XGBoost algorithm are, e.g., the learning rate, max. tree depth, and regularization parameters.

\subsection{Score Predictor Inference}
For inference, the trained predictors are integrated into the execution pipeline, as shown in \cref{fig:toolworkflow}.
A challenge arises because input parameters, such as $P_{L1Dw,norm}(I_x)$, cannot be determined for new, unknown groups.
This limitation is due to the Auto-Scheduler generating implementations batch-wise based on prior scores (see \cref{sec:autotuning}),
preventing the computation of mean values like ${\overline{P_{L1Dw}(\mathbf{I})}}$ at the beginning.

To address this, we allow approximating mean values using two approaches:
\textit{static} and \textit{dynamic} windows.
In the static approach, mean values are calculated from the first ${\overline{w}}$ samples.
In the dynamic window approach, mean values are adaptively adjusted over time.
The batch size, and thus the window size ${\overline{w}}$, is typically large enough that no accuracy loss compared to using 
${\overline{P_{L1Dw}(\mathbf{I})}}$  was observed in the experiments.

%% file: results.tex
\section{Results}
\label{sec:results}

To verify the quality of our predictors, all benchmarks are executed on three different CPU architectures: x86, ARM, and RISC-V.
For x86, we use a \SI{64}{\bit} \SI{2.2}{\giga\hertz} AMD Ryzen 7 5800X 8-Core processor;
for ARM, we use a \SI{64}{\bit} \SI{1.5}{\giga\hertz} Raspberry Pi 4 Model B with an ARM Cortex-A72 processor;
and for RISC-V, we use a \SI{64}{\bit} \SI{1.2}{\giga\hertz} SiFive U74-MC processor.

\cref{tab:cpus} lists the cache hierarchies of these CPUs.
We model them in gem5 (see \cref{sec:gem5setup}).
All cache line sizes are \SI{64}{\byte}.
The ARM and RISC-V \acp{cpu} feature  a shared L2 cache but no L3 cache.
All gem5 simulations are executed on the x86 machine.
We use five groups of Conv2D+Bias+ReLU kernels from a ResNet~\cite{wu2019wider} architecture as benchmarks.
\cref{tab:benchmarks} lists the shapes and parameters of these groups.
For training the predictors, the Auto-Scheduler generates \num{500} implementations per group, with \num{100} implementations used for the test set.

To determine the reference execution time ${t_{ref}}$, each implementation is executed ${N_{exe}=15}$ times, with the median value taken as the reference.
Additionally, cooldown times of ${t_{cooldown}=1s}$ are inserted between each run to ensure more reproducible measurements.
Workloads are not executed in parallel on real hardware, as this could affect the measurements.
This means that execution on ${K}$ parallel simulators is faster than native (sequential) execution on a single device.
\begin{align}
K = \left\lceil \frac{t_{simulator}}{\left(t_{cooldown} + t_{ref}\right) \cdot N_{exe}} \right\rceil 
\end{align}

Our measurement setup (${N_{exe}=15}$, ${t_{cooldown}=1s}$) results in ${K_{x86}\in[7,97]}$,
${K_{ARM}\in[4,31]}$, and ${K_{RISC-V}\in[3,21]}$ for the tested workloads.
This means that in the best case, \textit{only \num{3} parallel executions on the x86 machine are sufficient to achieve a speedup over native execution on the RISC-V CPU}.

\setlength{\tabcolsep}{2.1pt}

\begin{table}[]
    \vspace{0.2cm}
    \caption{Cache sizes and hierarchy of the used \acp{cpu}}
    \vspace{-0.1cm}
    \begin{tabular}{c|ccc|ccc|ccc|ccc}
    \multirow{2}{*}{} & \multicolumn{3}{c|}{L1 Data} & \multicolumn{3}{c|}{L1 Instruction} & \multicolumn{3}{c|}{L2} & \multicolumn{3}{c}{LLC (L3)} \\ \cline{2-13} 
     & \rotatebox{90}{size} & \rotatebox{90}{sets} & \rotatebox{90}{assoc } & \rotatebox{90}{size} & \rotatebox{90}{sets} & \rotatebox{90}{assoc} & \rotatebox{90}{size} & \rotatebox{90}{sets} & \rotatebox{90}{assoc} & \rotatebox{90}{size} & \rotatebox{90}{sets} & \rotatebox{90}{assoc} \\ \hline
    \rowcolor{black!15}x86    & 32K & 64 & 8    & 32K & 64 & 8     & 512K & 1024 & 8       & 32768K & 32768 & 16 \\
    ARM    & 32K & 256 & 2   & 48K & 256 & 3    & 1024K & 1024 & 16     & - & - & - \\
    \rowcolor{black!15}RISC-V & 32K & 64 & 8    & 32K & 64 & 8     & 2048K & 2048 & 16     & - & - & -      
    \end{tabular}
    \label{tab:cpus}
\end{table}

\setlength{\tabcolsep}{5pt}
\begin{table}[]
    \caption{Shapes of the used Conv2D+Bias+ReLU kernels}
    \vspace{-0.1cm}
    \centering
    \begin{tabular}{c|ccccccccc}
    group & N & H & W & CO & CI & KH & KW & stride & pad \\ \hline
    \rowcolor{black!15}0  & 1 & 224 & 224 &  64 &   3 & 7 &  7 & (2,2) &  (3,3)  \\
    1  & 1 &  56 &  56 &  64 &  64 & 3 &  3 & (1,1) &  (1,1)  \\
    \rowcolor{black!15}2  & 1 &  56 &  56 & 128 &  64 & 3 &  3 & (2,2) &  (1,1)  \\
    3  & 1 &  28 &  28 & 256 & 128 & 3 &  3 & (2,2) &  (1,1)  \\
    \rowcolor{black!15}4  & 1 &  14 &  24 & 512 & 256 & 3 &  3 & (2,2) &  (1,1) 
    \end{tabular}
    \label{tab:benchmarks}
    \vspace{-0.3cm}
\end{table}

\subsection{Prediction of Non-Trained Groups}
First, we demonstrate that a trained predictor can be utilized even when a specific kernel group was not included in the training data.
This capability is crucial as we aim to develop one predictor per architecture and kernel type (e.g., Conv2D+Bias+ReLU) that remains valid across all groups with different shapes and parameters (see \cref{sec:notations}).

\begin{figure*}[t!]
    \centering
   \begin{subfigure}[b]{0.3\linewidth}
       \centering
       \includegraphics[width=\linewidth, keepaspectratio]{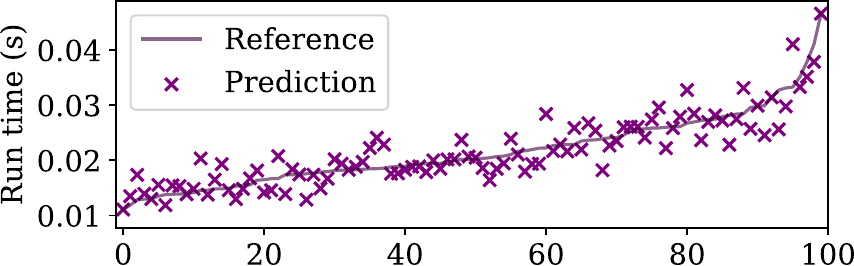}
       \caption{x86}
       \label{fig:x86gr1}
   \end{subfigure}
   \begin{subfigure}[b]{0.3\linewidth}
       \centering
       \includegraphics[width=\linewidth, keepaspectratio]{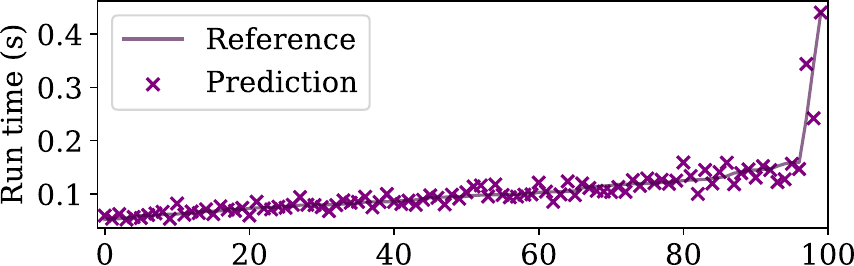}
       \caption{ARM}
       \label{fig:ARMgr1}
   \end{subfigure}
   \begin{subfigure}[b]{0.3\linewidth}
        \centering
        \includegraphics[width=\linewidth, keepaspectratio]{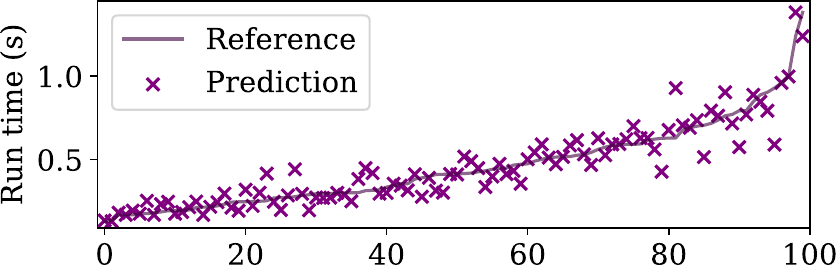}
        \caption{RISC-V}
        \label{fig:RISCVgr1}
    \end{subfigure}
    \\
    \vspace{0.2cm}
    \begin{subfigure}[b]{0.3\linewidth}
        \centering
        \includegraphics[width=\linewidth, keepaspectratio]{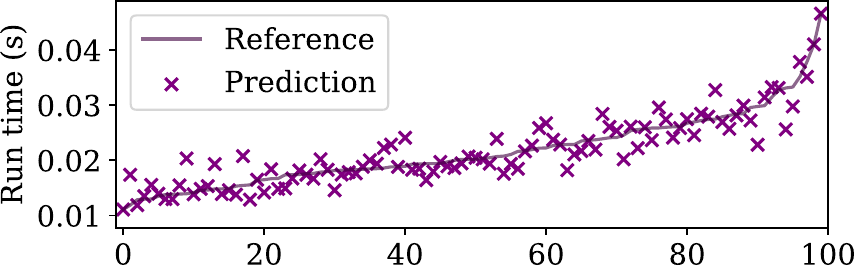}
        \caption{x86}
        \label{fig:x86gr3}
    \end{subfigure}
    \begin{subfigure}[b]{0.3\linewidth}
        \centering
        \includegraphics[width=\linewidth, keepaspectratio]{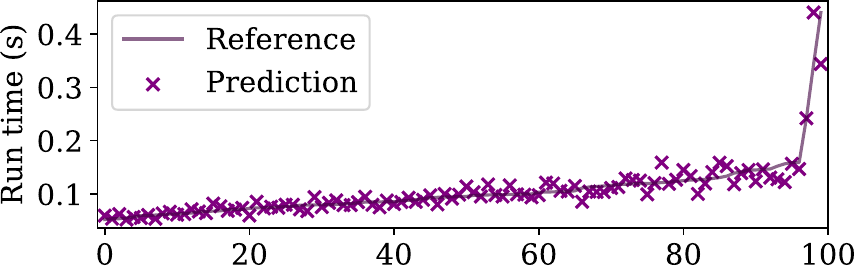}
        \caption{ARM}
        \label{fig:ARMgr3}
    \end{subfigure}
    \begin{subfigure}[b]{0.3\linewidth}
        \centering
        \includegraphics[width=\linewidth, keepaspectratio]{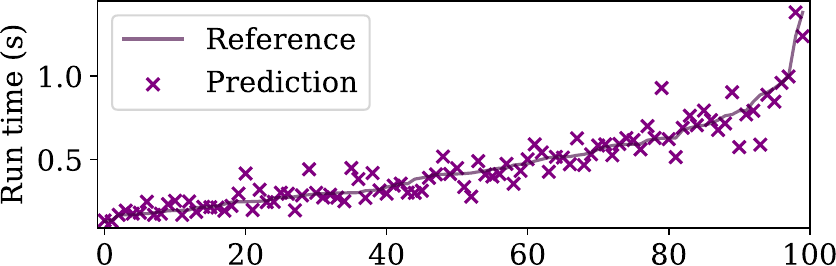}
        \caption{RISC-V}
        \label{fig:RISCVgr3}
    \end{subfigure}
    \caption{Sorted run time predictions for the test set of group 3 (a)-(c) when group 3 is \textbf{included} in the training vs. the same samples of group 3 (d)-(f) when group 3 is \textbf{not included} in the training}
    \label{fig:results}
    \vspace{-0.2cm}
\end{figure*}

To achieve this, we initially train Bayesian predictors for all architectures using all groups.
Subsequently, we train additional predictors using only groups 0, 1, 2, 4, and 5.
\cref{fig:results} compares the test set of group 3 when group 3 is included in the training (\cref{fig:x86gr1,fig:ARMgr1,fig:RISCVgr1}) against using the same samples when group 3 is not included in the training (\cref{fig:x86gr3,fig:ARMgr3,fig:RISCVgr3}).
The ${x}$-axis represents the individual samples.
The reference time ${t_{ref}}$ shows the median values of the measured, in ascending order sorted run times of the implementations.
To obtain the predicted run time ${t_{pred}}$, the predicted scores are sorted in ascending order,
with the corresponding measured run time plotted according to these scores.
A perfect prediction would mean that ${t_{pred}==t_{ref}}$.

Although not all scores are perfectly ordered, a clear ascending trend is evident.
Predictions for ARM and RISC-V architectures appear more accurate than those for x86.
This discrepancy may arise from fewer HW optimizations on these embedded \acp{cpu}.
Furthermore, each predictor is only as good as its reference measurements.
Since execution times on x86 are significantly faster, reference measurements show greater variability compared to longer run times on the embedded architectures.
Overall, visual inspections reveal no clear advantage between included and non-included training groups.
\textit{This demonstrates that the predictor can effectively be used even for groups that are not present in the training data.}
To enable better comparisons of predictions, we will introduce three different evaluation metrics in the following.

\subsection{Evaluation Metrics for the Predictors}
The most important aspect is predicting the fastest run time.
Therefore, we introduce ${E_{top1}}$ and the rank metric ${R_{top1}}$.
${E_{top1}}$ represents the relative error between the run times of the fastest reference measurement and the sample with the best predicted score:
\vspace{-0.4cm}
\begin{align}
    E_{top1} \coloneqq \left(1 - \frac{t_{ref}[0]}{t_{pred}[0]} \right) \cdot 100\%
\end{align}

Rank ${R_{top1}}$ indicates the relative position at which the fastest sample was ranked by the predictor:
\begin{align}
    R_{top1} \coloneqq \frac{100\%}{\lvert t_{ref} \rvert} \cdot \left( \underset{x}{\operatorname{argmin}} \left( t_{pred}[x] == t_{ref}[0] \right) + 1 \right)
\end{align}
For example, ${R_{top1}=}$~\SI{3}{\percent} means that the fastest sample was ranked within the top \SI{3}{\percent} of predictions.

Additionally, we need a metric to evaluate sorting quality.
Consecutive non-monotonically increasing samples should be penalized, as well as their extent of deviation.
Therefore, we introduce the Quality Score ${Q}$:
\begin{align}
    Q \coloneqq \frac{100\%}{\lvert t_{ref} \rvert} \cdot \sum_{i} \frac{t_{ref}[i] - min(t_{ref}[i], t_{ref}[i+1])} {t_{ref}[i]} 
\end{align}

To prevent deviations in the slower part of the run times from being disproportionately weighted, we evaluate ${Q}$ separately
for the lower \SI{50}{\percent} of the run times (${Q_{low}}$) and the higher \SI{50}{\percent} of the run times (${Q_{high}}$).
For all metrics, a smaller value indicates better performance.

\subsection{Comparison of Predictors}
Next, we compare the different predictors (see \cref{sec:predictortraining}).
We tested various loss functions, activation functions, and parameters.
Given that the XGBoost algorithm has many hyperparameters, we employed grid search~\cite{putatunda2018comparative} for tuning.
The used configurations (after tuning) are:

\begin{itemize}
\item \textbf{Linear Regression}: \ac{rss} loss
\item \textbf{\ac{dnn}}: 6 dense layers (number of neurons: 128, 128, 64, 32, 16, 1), linear output activation, tanh hidden layer activation, \ac{mae} loss, adam optimizer
\item \textbf{Bayes}: Objective function shown in \cref{lst:bayesFunction}, \ac{mse} loss
\item \textbf{XGBoost}: Column subsample ratio 0.6, learning rate 0.05, max. tree depth 3, alpha 0, lambda 0.1, gradient boost trees 300, min. child weight 1, subsample ratio for training 0.8, \ac{mse} loss
\end{itemize}

For the evaluation, all groups were included in the training.
One predictor (i.e., one of LinReg, DNN, etc.) was trained per \ac{cpu} architecture.
Each training was performed \num{10} times with a random selection of training and test sets.
Scores were subsequently calculated based on the median predictions from the test sets.
\cref{tab:x86results} shows the results for x86, \cref{tab:ARMresults} for ARM, and \cref{tab:Riscvresults} for RISC-V.

\setlength{\tabcolsep}{1.5pt}
\begin{table}[t]
\vspace{-0.2cm}
\centering
\caption{Prediction results for x86-based CPU}
\vspace{0.1cm}
\begin{tabular}{c|rrrr|rrrr|rrrr|rrrr}
\multirow{2}{*}{ID} & \multicolumn{4}{c|}{LinReg} & \multicolumn{4}{c|}{DNN} & \multicolumn{4}{c|}{Bayes} & \multicolumn{4}{c}{XGBoost} \\ \cline{2-17} 
& \rotatebox{90}{${E_{top1}(\%)}$} & \rotatebox{90}{${Q_{low}(\%)}$} & \rotatebox{90}{${Q_{high}(\%)}$ } & \rotatebox{90}{${R_{top1}(\%)}$}
& \rotatebox{90}{${E_{top1}(\%)}$} & \rotatebox{90}{${Q_{low}(\%)}$} & \rotatebox{90}{${Q_{high}(\%)}$} & \rotatebox{90}{${R_{top1}(\%)}$}
& \rotatebox{90}{${E_{top1}(\%)}$} & \rotatebox{90}{${Q_{low}(\%)}$} & \rotatebox{90}{${Q_{high}(\%)}$} & \rotatebox{90}{${R_{top1}(\%)}$}
& \rotatebox{90}{${E_{top1}(\%)}$} & \rotatebox{90}{${Q_{low}(\%)}$} & \rotatebox{90}{${Q_{high}(\%)}$} & \rotatebox{90}{${R_{top1}(\%)}$}
    \\ \hline
\rowcolor{black!15}0&10.7&3.0&3.0&8.5&1.4&2.7&2.2&3.0&12.5&3.0&2.7&3.0&7.0&2.7&2.4&2.0
\\
1&11.2&3.6&3.5&3.0&3.2&2.8&2.8&2.0&8.9&2.6&2.8&3.5&1.5&2.7&2.4&2.5
\\
\rowcolor{black!15}2&15.0&3.2&3.2&3.0&7.6&2.7&2.3&2.0&0.0&2.6&2.3&1.0&2.2&2.4&2.2&2.0
\\
3&8.8&3.2&3.0&2.0&0.7&2.4&2.4&1.5&0.7&2.7&2.7&1.5&0.0&2.8&2.4&1.0
\\
\rowcolor{black!15}4&0.0&3.4&3.3&1.0&2.8&3.1&2.2&3.0&5.4&2.5&2.6&2.0&2.0&2.8&2.4&2.0
\\
\end{tabular}
\label{tab:x86results}
\vspace{-0.5cm}
\end{table}

For the x86 architecture, the \ac{dnn}, Bayesian optimization, and XGBoost yield good scores.
XGBoost achieves the best average ${R_{top1}}$ score, with a maximum of \SI{2.5}{\percent}.
XGBoost also had the smallest prediction error, with an average of ${E_{top1}=}$ \SI{2.54}{\percent}.
For ARM and RISC-V, the scores are slightly better.
For ARM, DNNs, Bayesian optimization, and XGBoost achieve a ${R_{top1}}$ score smaller or equals \SI{2.5}{\percent}.
All ${E_{top1}}$ errors are below \SI{5}{\percent}, and often even below \SI{2}{\percent}.
Bayesian optimization correctly predicts the optimum in three out of five cases
and demonstrates the lowest ${Q_{low}}$ and ${Q_{high}}$.
For RISC-V, even the linear model performs quite well, with exceptions in ${E_{top1}}$ for group 0 and 4.
For other predictors, the optimum is often included in the top \SI{2}{\percent} of predictions.
The \ac{dnn} provides the best results with ${E_{top1}} \leq$~\SI{3.6}{\percent} and ${R_{top1}}\leq$~\SI{2}{\percent}.

\setlength{\tabcolsep}{1.5pt}
\begin{table}[h]
\centering
\vspace{-0.2cm}
\caption{Prediction results for ARM-based CPU}
\vspace{0.1cm}
\begin{tabular}{c|rrrr|rrrr|rrrr|rrrr}
\multirow{2}{*}{ID} & \multicolumn{4}{c|}{LinReg} & \multicolumn{4}{c|}{DNN} & \multicolumn{4}{c|}{Bayes} & \multicolumn{4}{c}{XGBoost} \\ \cline{2-17} 
& \rotatebox{90}{${E_{top1}(\%)}$} & \rotatebox{90}{${Q_{low}(\%)}$} & \rotatebox{90}{${Q_{high}(\%)}$ } & \rotatebox{90}{${R_{top1}(\%)}$}
& \rotatebox{90}{${E_{top1}(\%)}$} & \rotatebox{90}{${Q_{low}(\%)}$} & \rotatebox{90}{${Q_{high}(\%)}$} & \rotatebox{90}{${R_{top1}(\%)}$}
& \rotatebox{90}{${E_{top1}(\%)}$} & \rotatebox{90}{${Q_{low}(\%)}$} & \rotatebox{90}{${Q_{high}(\%)}$} & \rotatebox{90}{${R_{top1}(\%)}$}
& \rotatebox{90}{${E_{top1}(\%)}$} & \rotatebox{90}{${Q_{low}(\%)}$} & \rotatebox{90}{${Q_{high}(\%)}$} & \rotatebox{90}{${R_{top1}(\%)}$}
    \\ \hline

\rowcolor{black!15}0&9.9&3.4&2.9&3.5&0.2&2.8&2.3&1.5&0.0&2.6&2.0&1.0&2.7&3.0&2.2&1.5
\\
1&6.4&3.6&3.2&3.0&4.6&3.2&2.6&2.0&4.6&3.4&2.4&2.0&4.3&3.2&2.6&2.5
\\
\rowcolor{black!15}2&7.7&3.6&2.3&5.0&0.7&3.3&2.4&1.5&4.3&3.2&2.3&2.5&0.3&3.4&2.2&1.5
\\
3&7.7&2.8&2.6&2.5&1.1&2.8&2.3&1.5&0.0&2.8&2.0&1.0&4.0&3.2&2.2&2.0
\\
\rowcolor{black!15}4&2.2&3.5&2.5&4.0&0.2&3.1&2.6&1.5&0.0&2.6&2.2&1.0&1.0&3.0&2.3&2.0
\\
\end{tabular}
\label{tab:ARMresults}
\end{table}

\setlength{\tabcolsep}{1.5pt}
\begin{table}[h]
\vspace{0.2cm}
\caption{Prediction results for RISC-V-based CPU}
\vspace{0.2cm}
\centering
\begin{tabular}{c|rrrr|rrrr|rrrr|rrrr}
\multirow{2}{*}{ID} & \multicolumn{4}{c|}{LinReg} & \multicolumn{4}{c|}{DNN} & \multicolumn{4}{c|}{Bayes} & \multicolumn{4}{c}{XGBoost} \\ \cline{2-17} 
& \rotatebox{90}{${E_{top1}(\%)}$} & \rotatebox{90}{${Q_{low}(\%)}$} & \rotatebox{90}{${Q_{high}(\%)}$ } & \rotatebox{90}{${R_{top1}(\%)}$}
& \rotatebox{90}{${E_{top1}(\%)}$} & \rotatebox{90}{${Q_{low}(\%)}$} & \rotatebox{90}{${Q_{high}(\%)}$} & \rotatebox{90}{${R_{top1}(\%)}$}
& \rotatebox{90}{${E_{top1}(\%)}$} & \rotatebox{90}{${Q_{low}(\%)}$} & \rotatebox{90}{${Q_{high}(\%)}$} & \rotatebox{90}{${R_{top1}(\%)}$}
& \rotatebox{90}{${E_{top1}(\%)}$} & \rotatebox{90}{${Q_{low}(\%)}$} & \rotatebox{90}{${Q_{high}(\%)}$} & \rotatebox{90}{${R_{top1}(\%)}$}
    \\ \hline

\rowcolor{black!15}0&10.9&4.0&3.9&4.0&2.2&4.0&4.0&2.0&0.0&3.8&4.2&1.0&0.0&3.4&4.1&1.0
\\
1&0.0&4.0&4.3&1.0&0.6&3.7&4.5&1.5&2.5&3.4&4.4&2.0&4.6&3.5&4.5&2.5
\\
\rowcolor{black!15}2&0.0&3.4&3.7&1.0&0.0&3.2&3.8&1.0&0.0&2.8&3.4&1.0&0.0&3.0&3.8&1.0
\\
3&0.0&3.6&3.8&1.0&0.0&3.2&3.9&1.0&2.0&2.8&3.7&1.5&4.4&3.0&3.6&2.0
\\
\rowcolor{black!15}4&11.0&4.0&4.2&3.0&3.6&3.6&4.2&1.5&10.7&3.2&4.0&2.0&8.2&3.8&4.0&3.0
\\
\end{tabular}
\label{tab:Riscvresults}
\vspace{-0.3cm}
\end{table}

In summary, it is possible to train predictors for forecasting.
\textit{For RISC-V and ARM, a prediction error $E_{top1}$ of less than \SI{5}{\percent} is achievable.}
If the goal is to find the best sample, it is \textit{sufficient to re-execute the top \SI{2}{\percent}-\SI{3}{\percent} of the predictions} later on a real architecture.